\documentclass{moriond}

\usepackage{amsmath}
\usepackage{booktabs}

\def\Journal#1#2#3#4{{#1} {\bf #2}, #3 (#4)}

\def\PRL{\em Phys. Rev. Lett.}
\def\PRD{{\em Phys. Rev.} D}

\def\be{\begin{equation}}
\def\ee{\end{equation}}
\def\bea{\begin{eqnarray}}
\def\eea{\end{eqnarray}}

\newcommand{\Photo}{}

\begin{document}
\vspace*{4cm}
\title{Constraints on Lorentz symmetry violations with Lunar Laser Ranging.}

\author{A. \textsc{Bourgoin}}
\address{Dipartimento di Ingegneria Industriale, University of Bologna, via fontanella 40, Forl\`i,  Italy}

\author{A. \textsc{Hees}}
\address{Department of Physics and Astronomy, University of California, Los Angeles, California 90095, USA}

\author{C. \textsc{Le Poncin-Lafitte}, S. \textsc{Bouquillon}, G. \textsc{Francou}, and M.-C. \textsc{Angonin}}
\address{SYRTE, Observatoire de Paris, PSL Research University, CNRS, Sorbonne Universit\'es, UPMC Univ. Paris 06, LNE, 61 avenue de l'Observatoire, 75014 Paris, France}

\maketitle\abstracts{
We present new constraints on Lorentz symmetry (LS) violations with lunar laser ranging (LLR). Those constraints are derived in the standard-model extension (SME) framework aiming at parameterizing any LS deviations in all sectors of physics. We restrict ourself to two sectors namely the pure gravitational sector of the minimal SME and the gravity-matter coupling. We describe the adopted method and compare our results to previous analysis based on theoretical grounds. This work constitutes the first direct experimental determination of the SME coefficients using LLR measurements.}

\section{Introduction}

General relativity (GR) and the standard-model of particle physics provide a comprehensive description of nature. On one hand, GR describes gravitational effects as the classical consequence of space-time curvature by its energy content. On the other hand, the standard-model of particles describes all other non-gravitational forces by the quantum exchange of subatomic particles. They both assume a unique symmetry of space-time known as the Lorentz-symmetry (LS). A formulation of a quantum theory of gravity would be useful at the Planck era or for the study of black hole's singularities where quantum effects are supposed to become relevant. In many scenarios such a new physics would be expected to break some fundamental symmetry of space-time like the LS. An effective field theory built to consider all hypothetical LS violations in all sectors of physics is an efficient tool \cite{2004PhRvD..69j5009K} to classify and enumerate LS breaking. This effective field theory is currently known as the standard-model extension (SME) framework and proposes many parametrization of LS violations in many sectors \cite{KR2011}. In that work, we focused on two sectors namely the pure gravitational sector \cite{BK2006} and the classical point-mass limit in the matter sector~\cite{KT2011} of the SME. The main idea is to constrain the SME coefficients by analyzing lunar laser ranging (LLR) observations. In the literature, there already exist some evaluation of the SME coefficients in the pure gravitational sector using LLR observations \cite{BA2007}, however, it is based on theoretical grounds. Unfortunately, analysis based on theoretical grounds has been shown to be not fully satisfactory \cite{CLPLHL2016,BO2016} and to provide overoptimistic uncertainties. In order to provide real constraints, we have determined SME coefficient estimates directly from experimental measurements by performing a global LLR data processing directly in the SME framework \cite{BO2016}. To account for the effect of LS breaking upon the orbital motion of the Moon we have built a new lunar ephemeris named \emph{\'eph\'em\'eride lunaire parisienne num\'erique} (ELPN) computed directly in the SME framework. It integrates numerically the differential equations describing the orbital and the rotational motion of the Moon by taking into account all the theoretical effects producing a signal larger than the millimeter over the Earth-Moon distance \footnote{ the current observational accuracy being subcentimetric.}.

In section \ref{sec:GR} and \ref{sec:SME}, we describe respectively the solution obtained with ELPN in pure GR and in the SME framework. Finally, in section \ref{sec:jack} we discuss the method used to provide real constraints on the SME coefficients and we present new results obtained in both the pure gravitational sector of the minimal SME and the classical point-mass limit in the matter sector.

\section{Data analysis in pure general relativity}
\label{sec:GR}

\begin{figure}
  \begin{center}
    \includegraphics[scale=0.95]{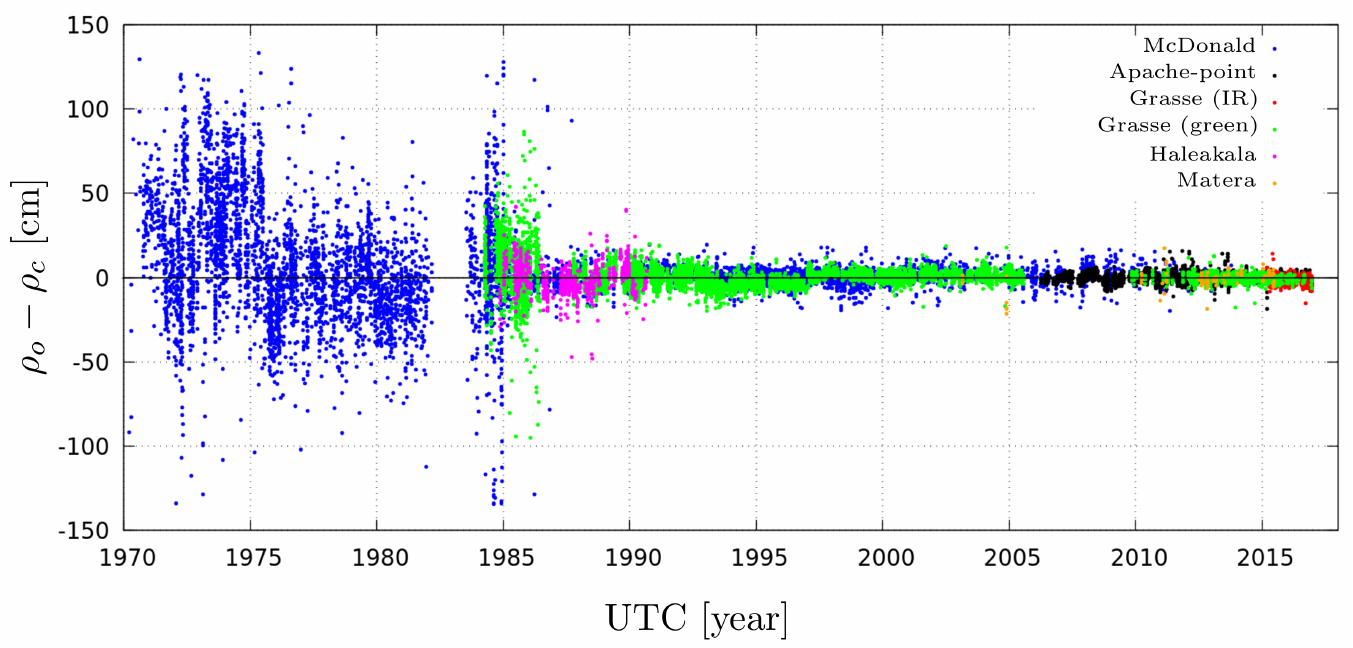}
  \end{center}
  \vspace{-0.3cm}
  \caption{Plot of the dispersion of the ELPN's residuals as a function of time per LLR stations. $\rho_c$ is the computed time delay determined with ELPN's predictions and $\rho_o$ is the observed time delay (normal points).}
  \label{fig:res}
\end{figure}

Analysis in the pure GR is a mandatory step which must be seen as a validation step to ensure that our ELPN solution produces residuals with a dispersion similar to the one currently obtained with other planetary and lunar ephemerides as DE430 \cite{FW2014} or INPOP13 \cite{FI2009}. To ensure that the precision of ELPN is at the same level of accuracy than other ephemerides, we have implemented the exact same equations of motion as in DE430 \footnote{ the main difference between INPOP13 and DE430 being the inclusion of a fluid lunar core in DE430.}. After integration, a direct comparison of ELPN predicted Earth-Moon distance with the one predicted by DE430 shows a maximum difference remaining below 4 cm during the time span of LLR observations. However, the true precision of our solution can only be determined after a fitting process to true LLR data by looking at the dispersion of the residuals. The residuals are obtained (i) by changing ELPN orbital predictions into observables i.e. into a prediction of the round trip light time at the date of each LLR normal points and (ii) by computing the difference between the observed light time and the computed one. The step (i) is achieved thanks to the 2010 international Earth rotation system conventions \cite{IERS}. The basic procedure for building the ELPN solution in pure GR can be described as follow. First, the physical parameters and the initial conditions are taken from DE430 to perform the first numerical integration. Then, the independent solution ELPN is built at the end of an iterative process consisting of adjusting 59 parameters \footnote{ like the position of the LLR stations and retroreflectors at J2000, the orbital and rotational lunar initials conditions at J2000, the masses of the Moon and the Earth-Moon barycenter, the Love's numbers and the time delays of degree 2 for solid body tides of both the Earth and the Moon, the total moment of inertia of the Moon, and the damping term between the mantle and the fluid core of the Moon.} to the LLR normal points and reintegrating the solution with the new set of constants. During this process 24022 normal points are considered spanning 48 years of LLR observations from August 1969 to December 2016. Among these normal points, we have considered 1337 observations in infrared wavelength obtained at Grasse LLR station in France. The final residuals obtained per LLR stations are shown in Fig. \ref{fig:res}. For the most recent observations (after 2006) the dispersion of the residuals is found to be 2.8 cm for Apache-point observatory in New Mexico (United states) with no offset on the mean, and respectively 2.4 cm and 2.2 cm for the Grasse station respectively in green and infrared wavelength. For these two last ones the mean of the dispersion is found to be respectively null and equals to -1.5 cm. This offset for the infrared wavelength has not yet been fully investigated and may correspond to (i) a bad position of the Grasse station in the international terrestrial reference frame for the LLR time span (non-constant tectonic drift), or (ii) a bad modeling for the tropospheric time delay correction in infrared wavelength, or (iii) a bad experimental calibration for that wavelength. The proposition (i) seems to be the most likely, but a more rigorous investigation is mandatory. Nevertheless, a comparison of those dispersions with the one obtained e.g. with INPOP13b \cite{FI2014} (see Tabs. 15 and 16) shows that our residuals are typically of the same order of magnitude and even better for the most recent observations. This lunar ephemeris obtained in pure GR will be used as a starting point for considering violations of LS in order to provide real constraints on GR violations. 

\section{Data analysis in the standard-model extension framework}
\label{sec:SME}

The most important difference between ELPN and DE430 or INPOP13 are the Lorentz violating contributions which are implemented in ELPN. The Lorentz violating terms are considered in two different parts of the modeling: (i) the orbital part, (ii) the gravitational delay for the light propagation. Each of these contributions can be split into two parts. The first one, which has already been discussed in a previous paper \cite{BO2016} comes from the pure gravitational sector of the minimal SME and is produced by LS violations at the level of the field equations. The second one comes from the classical point-mass limit in the matter sector of the SME framework and leads to violations at the level of the Einstein equivalence principle (EEP). The LS violations on the orbital motion of the Moon is characterized by supplementary accelerations computed in Eqs. (104) and (177) of respectively \cite{BK2006} and \cite{KT2011}. On the other hand, we have also considered the impact of LS violations on the gravitational time delay for light propagation computed by Eqs. (24) and (203) from respectively \cite{BQ2009} and \cite{KT2011}. Starting from the ELPN solution computed in pure GR (cf. Sec. \ref{sec:GR}), we have have produced a new ephemeris by fitting the SME coefficients in addition to the same parameters as considered previously. Some correlations between SME coefficients are very high meaning that the LLR data does not allow to estimate all the SME coefficients independently but only some combinations of them. An iterative investigation of partial derivatives let to assess the 6 most sensitive following independent linear combinations
\begin{subequations}\label{eq:comb}
  \begin{align}
    \bar s^1&=\bar s^{XY}\label{eq:combXY}\\
    \bar s^2&=\bar s^{YZ}\label{eq:combYZ}\\
    \bar s^3&=\bar s^{XX}-\bar s^{YY}\text{,}\label{eq:combA}\\
    \bar s^4&=0.35\bar s^{XX}+0.35\bar s^{YY}-0.70\bar s^{ZZ}-0.94\bar s^{YZ}\text{,}\allowdisplaybreaks\label{eq:combD}\\
    \bar s^5&=-0.62\bar s^{TX}+0.78\alpha(\bar a_{\text{eff}}^{e+p})^X+0.79\alpha(\bar a_{\text{eff}}^{n})^X\text{,}\label{eq:comb1}\\
    \bar s^6&=0.93\bar s^{TY}+0.33\bar s^{TZ}-0.10\alpha(\bar a_{\text{eff}}^{e+p})^Y-0.10\alpha(\bar a_{\text{eff}}^{n})^Y-0.044\alpha(\bar a_{\text{eff}}^{e+p})^Z-0.044\alpha(\bar a_{\text{eff}}^{n})^Z\text{.}\label{eq:comb2}
  \end{align}
\end{subequations}
Interestingly, previous work based on theoretical grounds \cite{AH2015} has also shown that LLR is sensitive to 6 linear combinations of the fundamental SME coefficients. Nevertheless, a look at Eqs. \eqref{eq:comb} shows that the extension to the matter sector only change 2 linear combinations compared to previous global LLR data analysis \cite{BO2016} in the pure gravitational sector of the minimal SME. On the opposite, theoretical work \cite{AH2015} expects the modification of 4 linear combinations. This determination of the linear combinations that can be constrained by direct experimental measurements highlights the limit of analyzes that rely on theoretical grounds and the need to consider a full reduction of the raw data within the SME formalism.

\section{Constraints on the SME coefficients}
\label{sec:jack}

The SME coefficients which can be constrained with LLR data are given by Eqs. \eqref{eq:comb}. An adjustment of these coefficients together with the other global parameters provide an estimate of the SME coefficients as well as their statistical marginalized uncertainties. Unlike constraints derived from a postfit analysis, these uncertainties take into account the correlations with the other global parameters. However, considering subsets of our dataset reveals that the estimates of the SME coefficients depend highly on the LLR stations or retroreflectors. This is interpreted as systematics effects in our analysis which have been quantified by using a jackknife resampling method as discussed in \cite{BO2016}. The final estimates of SME coefficients including both statistical and systematics uncertainties are shown in Tab. \ref{tab:est}.

\begin{table}
  \begin{center}
    {\normalsize
    \begin{tabular}{|c|l|}
      \hline
      SME & \multicolumn{1}{|c|}{constraints} \\
      \hline
      $\bar s^1$ & $(-0.5\pm3.6)\times10^{-12}$ \\
      $\bar s^2$ & $(+2.1\pm3.0)\times10^{-12}$ \\
      $\bar s^3$  & $(+0.2\pm1.1)\times10^{-11}$ \\
      $\bar s^4$  & $(+3.0\pm3.1)\times10^{-12}$ \\
      $\bar s^5$  & $(-1.4\pm1.7)\times10^{-8}$\\
      $\bar s^6$  & $(-6.6\pm9.4)\times10^{-9}$\\
      \hline
    \end{tabular}
    }
  \end{center}
  \caption{Real constraints on SME coefficients derived from a global LLR data analysis. The quoted uncertainties correspond to $1\sigma$ realistic uncertainties. The linear combinations are defined in Eqs. \eqref{eq:comb}.}
  \label{tab:est}
\end{table}

\section{Conclusion}

Our results do not show any deviations from GR at the 68\% confidence level. In addition, they constitute the first real constraints from a direct experimental measurement with LLR data in both the pure gravitational sector of the minimal SME and the classical point-mass limit in the matter sector. LLR data should be analyzed with other measurements in order to provide estimates for each individual SME coefficients \cite{AH2015}. 

\section*{Acknowledgments}

A.B. thanks SYRTE from Paris Observatory for financial support. A.B. and A.H. thank the organizers for financial support to attend the conference.


\section*{References}

\begin{thebibliography}{99}
\bibitem{2014LRR....17....4W} C. M. Will, {\em Living Reviews in Relativity}, {\textbf{10}}, 12942 (2014).
\bibitem{2004PhRvD..69j5009K} V. A. Kosteleck\'y, \Journal{\PRD}{69}{105009}{2004}.
\bibitem{KR2011} V. A. Kosteleck\'y and N. Russell, {\em Reviews of Modern Physics}, {\textbf{83}}, 11 (2011).
\bibitem{BK2006} Q. G. Bailey and V. A. Kosteleck\'y, \Journal{\PRD}{74}{045001}{2006}.
\bibitem{KT2011} V. A. Kosteleck\'y and J. D. Tasson, \Journal{\PRD}{83}{016013}{2011}.
\bibitem{BA2007} J. B. R. Battat \emph{et al.}, \Journal{\PRL}{99}{241103}{2007}.
\bibitem{CLPLHL2016} C. Le Poncin-Lafitte, \emph{et al.}, \Journal{\PRD}{94}{125030}{2016}.
\bibitem{BO2016} A. Bourgoin, \emph{et al.}, \Journal{\PRL}{117}{241301}{2016}.
\bibitem{FW2014} W. M. Folkner, \emph{et al.}, {\em Interplanetary Network Progress Report}, {\textbf{42-196}}, 1-81 (2014).
\bibitem{FI2009} A. Fienga, \emph{et al.}, {\em Astronomy and Astrophysics}, {\textbf{507}}, 1675-1686, (2009).
\bibitem{IERS} G. Petit \emph{et al.}, {\em IERS Technical Note}, {\textbf{36}}, (2010).
\bibitem{FI2014} A. Fienga, \emph{et al.}, arXiv:astro-ph.EP/1405.0484.
\bibitem{BQ2009} Q. G. Bailey, \Journal{\PRD}{80}{044004}{2009}.
\bibitem{AH2015} A. Hees, \emph{et al.}, \Journal{\PRD}{92}{064049}{2015}.
\end{thebibliography}
\end{document}